\documentstyle[prl,preprint,aps]{revtex}

\def\bra#1{\langle #1 |}
\def\ket#1{| #1 \rangle}
\def\bracket#1#2{\langle #1 | #2 \rangle}
\def\expect#1{\langle #1 \rangle}
\def\ident{{\hat 1}}
\def\H{{\hat H}}
\def\Ld{{\hat L}}
\def\Heff{{\H_{\rm eff}}}
\def\P{{\hat {\cal P}}}
\def\L{{\cal L}}
\def\e{{\rm e}}
\def\Tr{{\rm Tr}}
\def\a{{\hat a}}
\def\adag{{\hat a}^\dagger}
\def\b{{\hat b}}
\def\bdag{{\hat b}^\dagger}

\tighten
\begin{document}

\title{Quantum jumps as decoherent histories}

\author{Todd A. Brun\cite{Current} \\
Department of Physics, Queen Mary and Westfield College \\
University of London, London E1 4NS, England}

\date{\today}

\maketitle

\begin{abstract}
Quantum open systems are described in the Markovian limit by master
equations in Lindblad form.  I argue that common ``quantum jumps''
techniques, which solve the master equation
by unraveling its evolution into stochastic trajectories in Hilbert space,
correspond closely to a particular set of decoherent
histories.  This is
illustrated by a simple model of a photon counting experiment.
\end{abstract}

Recently a great deal of work has been done in
quantum optics on ``quantum jump'' simulations of continuously
measured systems with dissipation
\cite{Carmichael1,Dalibard,Gardiner,Wiseman1,Carmichael2}.
In this technique, a system
described by a master equation for the reduced density operator $\rho$
in the Markovian approximation \cite{Lindblad},
\begin{equation}
{\dot \rho} = - i [\H,\rho] + \sum_m \Ld_m \rho \Ld_m^\dagger
  - {1\over2} \Ld_m^\dagger \Ld_m \rho
  - {1\over2} \rho \Ld_m^\dagger \Ld_m,
\label{master_eqn}
\end{equation}
is ``unraveled'' into a jump process for pure states.
$\H$ is the system Hamiltonian, and the $\{\Ld_m\}$ are a set
of {\it Lindblad operators} which model the effects of the environment.

Around the same time,
the decoherent histories formulation of quantum mechanics was developed
\cite{Griffiths,Omnes1,Omnes2,GMHart1,GMHart2,Omnes3}.
In this formalism, one describes a quantum system in terms of an exhaustive
set of possible histories, which must satisfy a {\it decoherence}
or {\it consistency} criterion.
Histories which satisfy this criterion have
probabilities which obey the usual classical
probability sum rules.

Both quantum trajectories and decoherent histories describe a quantum system
in terms of alternative possible evolutions; they thus bear a certain
resemblance to each other.  What is more, quantum jumps are commonly
interpreted as giving the results of continuous measurements;
and histories which correspond
to records of a ``classical'' measuring device should
always decohere \cite{GMHart1}.  Thus, there should be a set of
decoherent histories which correspond to the quantum trajectories of a
continuously measured system.

Exactly such a correspondence has recently been shown between decoherent
histories and quantum state diffusion (QSD), another unraveling of
the master equation, by Di\'osi, Gisin, Halliwell and Percival \cite{DGHP}.
Though this result was pioneering, it was rather abstract, and lacked any
direct connection to a physical measurement situation.  Similar results
for yet another unraveling were given by Paz and Zurek \cite{PazZurek}
and Di\'osi \cite{Diosi1}, in a model with exact decoherence, but also
far removed from physical measurement situations.
Other treatments \cite{Nielsen} have been framed in terms
of measurement alone.

Consider a quantum system with a Hamiltonian $\H_0$,
completely isolated except for a single channel of decay, which is monitored
by an external photon detector.  We
model this detector as a single
two-level system (the ``output mode'') with states $\ket0$ and
$\ket1$ strongly coupled to an environment
representing the remaining degrees of freedom of the device.

The measuring device produces two important
effects.  The first is dissipation.  Excitations of
the output mode will be absorbed by the measuring device with a rate
$\Gamma_1$ which we assume to be rapid compared to the dynamical time-scale
of the system.  The time $1/\Gamma_1$ represents the time-resolution of
the detector.

The second effect is more subtle but just as important:  decoherence.  As
the state of the output mode becomes correlated with the internal
degrees of freedom of the measuring device, the phase coherence between
the ground and excited states of the output mode is lost.  Investigations
of this process have shown that the loss of coherence is generally far
quicker than the actual rate of energy loss \cite{Zurek}.
This decoherence rate is
$\Gamma_2 \gg \Gamma_1$.

We suppose that the system is linearly coupled to the output mode
via the Hamiltonian
\begin{equation}
\H_I = \kappa ( \adag \otimes \b + \a \otimes \bdag ),
\end{equation}
and the total Hamiltonian is
\begin{equation}
\H = \H_0 \otimes \ident
  + \kappa ( \adag \otimes \b + \a \otimes \bdag ),
\end{equation}
where $\a$ and $\b$ ($\adag$ and $\bdag$) are the lowering (raising)
operators for the system and output mode, respectively.
The hierarchy of evolution rates is $\Gamma_2 \gg \Gamma_1 \gg \kappa$.

The total system obeys the master equation
\begin{eqnarray}
{\dot\rho} && = - i [\H,\rho] + \Gamma_1 \b \rho \bdag
  - {\Gamma_1\over2} \bdag\b\rho
  - {\Gamma_1\over2} \rho\bdag\b \nonumber\\
&& + \Gamma_2 \sigma_z \rho \sigma_z - \Gamma_2 \rho
  \equiv {\cal L} \rho,
\label{total_master}
\end{eqnarray}
where $\rho$ is the density matrix for the combined system and output
mode, and the Pauli operator $\sigma_z$ acts on the output mode.
${\cal L}$ is the {\it Liouville superoperator}.
This is a linear equation, and so can be formally solved:
\begin{equation}
\rho(t_2) = \exp\biggl\{ {\cal L}(t_2 - t_1) \biggr\} \rho(t_1).
\label{liouville}
\end{equation}

Assume that we start in a pure state
$\ket\Psi = \ket\psi \otimes \ket0$.
We can expand $\rho$
\begin{equation}
\rho(t) = \rho_{00}(t) \otimes \ket0\bra0 + \rho_{01}(t) \otimes \ket0\bra1
  + \rho_{10}(t) \otimes \ket1\bra0 + \rho_{11}(t) \otimes \ket1\bra1.
\end{equation}
In terms of these components the master equation becomes
\begin{eqnarray}
{\dot\rho}_{00} = - i[\H_0,\rho_{00}] - i \kappa \adag \rho_{10}
  + i \kappa \rho_{01} \a + \Gamma_1 \rho_{11}, \nonumber\\
{\dot\rho}_{01} = - i[\H_0,\rho_{01}] - i \kappa \adag \rho_{11}
  + i \kappa \rho_{00} \adag - G \rho_{01}
  = {\dot\rho}_{10}^\dagger, \nonumber\\
{\dot\rho}_{11} = - i[\H_0,\rho_{11}] - i \kappa \a \rho_{01}
  + i \kappa \rho_{10} \adag - \Gamma_1 \rho_{11},
\label{components}
\end{eqnarray}
where $G = \Gamma_1/2 + 2\Gamma_2 \gg \Gamma_1 \gg \kappa$.
(This combination $G$ occurs frequently in the equations which follow.)

Since the $\rho_{01}, \rho_{10}, \rho_{11}$ components are heavily damped,
we can adiabatically eliminate
all components other than $\rho_{00}$ \cite{Wiseman2}:
\begin{equation}
{\dot\rho}_{00} = - i [\H_0,\rho_{00}]
  + {2\kappa^2\over G} \a \rho_{00} \adag
  - {\kappa^2\over G} \adag\a \rho_{00}
  - {\kappa^2\over G} \rho_{00} \adag\a,
\label{adiabatic}
\end{equation}
to first order in $\kappa^2/G$, provided that the system is not so highly
excited as to emit too rapidly, i.e.,
$\kappa \expect{\adag\a} \ll \Gamma_1$.

We can unravel the master equation (\ref{adiabatic}) into a sum over
quantum jump trajectories.  First, define a non-Hermitian
{\it effective Hamiltonian}
\begin{equation}
\H_{\rm eff} = \H_0 - i (\kappa^2/G) \adag\a.
\label{heff}
\end{equation}
Assume that the system (excluding the output mode)
begins in a pure state $\ket\psi$.  $\ket\psi$
evolves according to the Schr\"odinger equation,
\begin{equation}
{d\ket\psi\over dt} = - {i\over\hbar} \H_{\rm eff} \ket\psi,
\end{equation}
interrupted at random times by sudden quantum jumps
\begin{equation}
\ket\psi \rightarrow  \a\ket\psi.
\label{jump}
\end{equation}
These jumps correspond to the detection of photons
\cite{Carmichael1,Dalibard}.
Note that this evolution does not preserve the norm of the state.
The physical state is
taken to be $\ket{\tilde\psi} = \ket\psi/\sqrt{\bracket{\psi}{\psi}}$,
the renormalized state.

The probability that an initial state $\ket\psi$ evolves for a time $T$
and undergoes $N$ jumps during intervals $\delta t$ centered
at times $t_1, \ldots, t_N$ is
\begin{eqnarray}
&  (2\delta t\kappa^2/G)^N
  \Tr\biggl\{ \e^{-i\Heff(T-t_N)} \a \e^{-i\Heff(t_N - t_{N-1})} \a
  \cdots \a \e^{-i\Heff t_1} \nonumber\\
& \times \ket\psi\bra\psi \e^{i\Heff^\dagger t_1} \adag
  \cdots \adag \e^{i\Heff^\dagger(T-t_N)} \biggr\},
\label{jumps_prob}
\end{eqnarray}
i.e., the norm of the unrenormalized state gives the probability for that
state to be realized.

Equation (\ref{adiabatic}) is valid only as long as
the Markovian approximation remains good.  In the case of our toy model,
this means that it is valid only on time-scales longer than
$1/\Gamma_1$.  Thus, rather than a jump occurring at a time
$t_i$, it is more correct to consider the jump as occurring during an interval
$\delta t \sim 1/\Gamma_1$ centered on $t_i$.  This is fine so long as
the jumps are separated by more than $\delta t$ on average, i.e., the
system is not too highly excited.

By averaging $\ket{\tilde\psi}\bra{\tilde\psi}$ over all possible trajectories
with the probability measure (\ref{jumps_prob}),
one can show that this unraveling reproduces the master equation
(\ref{adiabatic}) as required \cite{Gardiner}.

Now, let us turn to the decoherent histories picture.
In non-relativistic quantum mechanics, a set of histories for a
system can be specified by choosing
a complete set of projections $\{\P^j_{\alpha_j}(t_j)\}$ at a
sequence of times $t_1,\ldots,t_N$,
which represent different exclusive alternatives:
\begin{equation}
\sum_{\alpha_j} \P^j_{\alpha_j}(t_j) = \ident,\ \ 
  \P^j_{\alpha_j}(t_j) \P^j_{\alpha_j'}(t_j) =
  \delta_{\alpha_j \alpha_j'} \P^j_{\alpha_j}(t_j).
\end{equation}

A particular history (denoted $h$)
is given by choosing one $\P$ at each point in time.
The {\it decoherence functional} on a pair of histories
$h$ and $h'$ is
\begin{equation}
D[h,h'] = \Tr \biggl\{ \P^N_{\alpha_N}(t_N) \cdots
  \P^1_{\alpha_1}(t_1) \rho(t_0) \P^1_{\alpha_1'}(t_1) \cdots
  \P^N_{\alpha_N'} \biggr\},
\label{functional}
\end{equation}
where $\rho(t_0)$ is the initial density matrix of the system
\cite{GMHart1}.  This satisfies the {\it decoherence criterion} if
the off-diagonal terms vanish, $D[h,h'] = 0, h \ne h'$.  The diagonal
terms then give the probabilities of the histories,  $p(h) = D[h,h]$.

Suppose our initial pure state is
$\ket\Psi = \ket{\psi_0} \otimes \ket0$,
and we consider histories composed only
of the Schr\"odinger projections
\begin{equation}
\P_0 = \ident \otimes \ket0\bra0,\ \ 
  \P_1 = \ident \otimes \ket1\bra1,
\end{equation}
representing the absence or presence of a photon in the
output mode.  The projections are spaced a short time $\delta t$
apart, and a history is composed of $N$ projections, representing a
total time $T = N\delta t$.  A single history $h$ is specified by a string
$\{\alpha_1,\alpha_2,\ldots,\alpha_N\}$, where $\alpha_j = 0,1$.
In this case, by the quantum regression theorem \cite{QRT} the decoherence
functional (\ref{functional}) becomes
\begin{equation}
D[h,h'] = \Tr \biggl\{ \P_{\alpha_N} \e^{\L\delta t}(
  \P_{\alpha_{N-1}} \e^{\L\delta t}( \cdots \e^{\L\delta t}(
  \P_{\alpha_1} \ket\Psi\bra\Psi \P_{\alpha_1'} )
  \cdots ) \P_{\alpha_N'} \biggr\}.
\label{jump_functional}
\end{equation}

The Liouville time evolution superoperators (\ref{liouville})
evolve pure states into mixed states.
This is counteracted by the effect of the
repeated projections $\P_\alpha$.

From the equations (\ref{components})
we can determine the character of the
different histories.  The crucial parameter is the size
of the spacing $\delta t$ between projections.
The interesting regime is in the range
\begin{equation}
{1\over G} \ll \delta t \ll {1\over\Gamma_1}.
\end{equation}
On this time-scale, the $\Gamma_2$ terms are sufficient to insure decoherence
while the effects of the $\Gamma_1$
terms are resolved into individual pure state trajectories.
This last is a subtle point.  The probability of a photon being emitted
in any single time step is small.  However, if a photon {\it is} emitted,
it has an appreciable possibility of being absorbed on a time scale
$1/\Gamma_1$.  The effect of decoherence produces
the terms $(\kappa^2/G)\adag\a\rho_{00}$ and $(\kappa^2/G)\rho_{00}\adag\a$
in equation (\ref{adiabatic}),
which are included in the effective Hamiltonian (\ref{heff}).  These terms
are already important on a time scale $\delta t \ll 1/\Gamma_1$. By contrast,
the term $(2\kappa^2/G)\a\rho_{00}\adag$ is produced by the effects of
dissipation, which only become important on a time scale $1/\Gamma_1$.
It is this term which causes pure states to evolve into mixed states
in equation (\ref{adiabatic}).  By choosing a time $\delta t \ll 1/\Gamma_1$,
we can maintain the purity of the system state over a full trajectory,
as we shall see.

If the external mode is initially unexcited, with
$\rho = \rho_{00} \otimes \ket0 \bra0$, then
after evolving for a time $\delta t$ the state becomes
\begin{eqnarray}
(\e^{\L\delta t}\rho)_{00} && = \rho_{00} - i [\H_0,\rho_{00}] \delta t
  - {\kappa^2\over G} \adag\a \rho_{00} \delta t
  - {\kappa^2\over G} \rho_{00} \adag\a \delta t + {\rm h.o.t.} \nonumber\\
&& \approx \e^{ - i (\H_0 - i(\kappa^2/G)\adag\a ) \delta t} \rho_{00}
  \e^{ i (\H_0 + i(\kappa^2/G)\adag\a ) \delta t}, \nonumber\\
(\e^{\L\delta t}\rho)_{01} && = {i\kappa\over G} \rho_{00} \adag
  + {\rm h.o.t.}
  = (\e^{\L\delta t}\rho)_{10}^\dagger, \nonumber\\
(\e^{\L\delta t}\rho)_{11} && = {2\kappa^2\over G} \a \rho_{00} \adag \delta t
  + {\rm h.o.t.}
\label{unexcited}
\end{eqnarray}
Here we see the appearance of the effective Hamiltonian $\Heff$, just
as in the quantum jump unraveling.

If the initial state is
$\rho = \rho_{11} \otimes \ket1 \bra1$,  after a time $\delta t$
the state becomes
\begin{eqnarray}
(\e^{\L\delta t}\rho)_{00} && = \Gamma_1 \delta t
  \e^{- i \Heff \delta t} \rho_{11} \e^{i \Heff^\dagger \delta t}
  + {2\kappa^2\over G} \adag \rho_{11} \a + {\rm h.o.t.} \nonumber\\
(\e^{\L\delta t}\rho)_{01} && = - {i\kappa\over G} \adag \rho_{11}
  + {\rm h.o.t.}
  = (\e^{\L\delta t}\rho)_{10}^\dagger, \nonumber\\
(\e^{\L\delta t}\rho)_{11}
&& = (1 - (\Gamma_1 + 2\kappa^2/G)\delta t)
  \e^{ - i \Heff \delta t} \rho_{11}
  \e^{ i \Heff^\dagger \delta t} + {\rm h.o.t.},
\label{excited}
\end{eqnarray}
Once again the effective Hamiltonian appears, together with two additional
effects.  The first is the possibility that the photon in the excited
mode will be absorbed by the measuring device.  The second (much smaller)
effect is the possibility that the photon will be coherently re-absorbed
by the system.  This last process is so weak as to be negligible.

By combining the above expressions with the appropriate
projections $\P_0$ and $\P_1$ (which pick out the $\rho_{00}$ or
$\rho_{11}$ component, respectively), we can write down the probabilities
of all possible histories.

Note that the magnitude of the off-diagonal $\rho_{01,10}$
terms in both cases is of
order $O(\kappa/G)$.  (This is also true for transitions from off-diagonal
to diagonal terms.)  This will be important in estimating the decoherence of
this set of histories.

First consider the history
given by an unbroken string of $N$ $\P_0$ projections, corresponding
to no photon being emitted during a time $N\delta t$.

The probability of such a history is given by the diagonal element
of (\ref{jump_functional}).  We can expand the time
evolution superoperator using (\ref{unexcited}) and see that
after $\delta t$ we get
\begin{equation}
\P_0 \e^{\L\delta t} (\ket{\psi}\bra{\psi} \otimes \ket0 \bra0) \P_0
  \approx \biggl(\e^{-i (\H_0 - i(\kappa^2/G) \adag\a) \delta t} \ket\psi
  \bra\psi \e^{i (\H_0 + i(\kappa^2/G) \adag\a) \delta t} \biggr)
  \ket0 \bra0.
\end{equation}
Repeating this $N$ times and taking the trace we get
\begin{equation}
p(h) \approx \Tr\biggl\{ \e^{- i \Heff N\delta t} \ket\psi
  \bra\psi \e^{i \Heff^\dagger N\delta t} \biggr\},
\label{no_jumps}
\end{equation}
which exactly agrees with the probability of the quantum jump trajectory
when no jumps are detected.

Suppose now that at time $N\delta t$ a photon is emitted, so that
instead of using a final projection $\P_0$ we use
$\P_1$.  This corresponds to keeping the $\rho_{11}$ component
of $\exp(\L\delta t)\rho$ instead of $\rho_{00}$,
and yields a probability
\begin{equation}
p(h) \approx (2\delta t\kappa^2/G) \Tr\biggl\{
  \a \e^{- i \Heff N\delta t} \ket\psi
  \bra\psi \e^{i \Heff^\dagger N\delta t} \adag \biggr\},
\label{one_jump}
\end{equation}

Once again, this exactly agrees with the probability of the corresponding
quantum jump trajectory.
What happens after the output mode has ``registered'' as being in the
excited state?  Essentially, there are two possibilities:  either the
output mode can drop back to the unexcited state (representing
absorption of the photon by the measuring device) or it can remain in
the excited state.
\begin{equation}
\P_0 \e^{\L\delta t} (\ket{\psi'} \bra{\psi'} \otimes \ket1 \bra1) \P_0
  \approx \Gamma_1 \delta t \ket{\psi'} \bra{\psi'} \otimes \ket0 \bra0,
\label{after_jump0}
\end{equation}
\begin{equation}
\P_1 \e^{\L\delta t} (\ket{\psi'} \bra{\psi'} \otimes \ket1 \bra1) \P_1
  \approx (1 - \Gamma_1 \delta t) \e^{-i\Heff\delta t} \ket{\psi'}
  \bra{\psi'} \e^{i\Heff^\dagger\delta t} \otimes \ket1\bra1.
\label{after_jump1}
\end{equation}

We see that the output mode has a probability of roughly
$\Gamma_1\delta t$ per time $\delta t$ of dropping back to the ground state,
while the system state continues to evolve according to the
effective Hamiltonian $\Heff$.

This is slightly different from quantum jumps.
Quantum jumps are resolved only on a time-scale $1/\Gamma_1$,
not $\delta t \ll 1/\Gamma_1$.  However, there is a near-unity
probability of the external mode returning to the ground state within a
time of order $1/\Gamma_1$, so one can simply sum over all
histories in which the photon is absorbed within this time.  It is easy to
see that these will, once again, match the quantum jump trajectories exactly.
This type of coarse-graining is common in decoherent histories
\cite{GMHart1,GMHart2}, and does not alter the form of the result.

By combining the three cases described in this section, one can produce
histories of multiple jumps.  It is clear that the probability of
such a history will be exactly of the form (\ref{jumps_prob}).

In order for this discussion of probabilities to be meaningful we must
require the histories to be decoherent.  Exact decoherence
is a very difficult criterion to meet.  It is
more usual to show that a model is {\it approximately} decoherent.
In order for the probability sum rules to
be satisfied to a precision $\epsilon \ll 1$ we require that \cite{DowkHall}
\begin{equation}
|D[h,h']|^2 < \epsilon^2 D[h,h] D[h',h'] = \epsilon^2 p(h) p(h'),
\end{equation}
for all distinct histories $h,h'$.  Generally speaking, the
more ``different'' a pair of histories is (i.e., the more projections
they differ in), the more suppressed the off-diagonal term.  So it
suffices to look at two histories which
differ at a single time $t_i$, one having a projection $\P_0$, the
other $\P_1$.  This is equivalent to
picking out the $\rho_{01}$ or $\rho_{10}$ component of
$\exp(\L\delta t)\ket{\psi'}\bra{\psi'}$ at that time.

Examining the components given by
(\ref{unexcited}--\ref{excited}),
\begin{equation}
{ |D[h,h']|^2 \over p(h) p(h') } \sim { 1 \over (G \delta t)^2 },
\end{equation}
we expect the sum rules to be obeyed with a precision of roughly
$O(1/G\delta t)$ (where we once again have assumed $\kappa \expect{\adag\a}$
is small compared with $\Gamma_1$).

We have seen how, in this simple model of a continuous measurement,
the set of quantum jump trajectories corresponds to
a set of decoherent histories.
One of the principal goals of the decoherent histories program
was to create a formalism which would reproduce the results of the
usual Copenhagen formalism in measurement situations.  It is pleasant
to note that extensions
to repeated or continuous measurements follow naturally within
decoherent histories.

In this letter, I considered only one measurement scheme:  direct
photodetection.  In fact, there are many different schemes which
give rise to different unravelings of the same master
equation---heterodyne and homodyne detection, to name two
\cite{Wiseman1,Diosi2}.  I have no doubt
that arguments similar to those I have advanced in this paper will
demonstrate similar correspondences to different
sets of decoherent histories.

This correspondence also has obvious practical benefits.
Enumerating a full set of decoherent histories and calculating their
probabilities is an arduous and unrewarding task, in general.
There is a great deal of
accumulated experience in simulating quantum trajectories;
in situations where one would like to generate individual
decoherent histories with correct probabilities, existing
numerical techniques could be used.

The decoherent histories formalism
was developed largely in response to the problems
of quantum cosmology, while quantum trajectories arose from
problems in quantum optics and atomic physics.
Both extend the von Neumann description
of quantum mechanics to new realms of application.
As the connections between
the two formalisms are further explored, we can hope that a great deal
of interesting physics will emerge.

I would like to thank Lajos Di\'osi, Murray Gell-Mann,
Nicolas Gisin, Jonathan Halliwell, Jim Hartle, Peter Knight, Mike
Nielsen, Ian Percival, Martin Plenio, and R\"udiger Schack for
valuable conversations, suggestions, and feedback.  After completion of
this research, I became aware of related work by Ting Yu, from a rather
different approach \cite{Yu}.  Financial support was provided by
the UK EPSRC.

\vfil
\end{document}